# Three faces of metabolites
## Pathways, localizations and network positions

**JING ZHAO**
Department of Mathematics, Logistical Engineering University, Chongqing 400016, China
Department of Natural Medicinal Chemistry, College of Pharmacy, Second Military Medical University, Shanghai, China
E-mail: zhaojanne@gmail.com, corresponding author

**PETTER HOLME**
Department of Physics, Umeå University, 90187 Umeå, Sweden
Computational Biology, Royal Institute of Technology, 10044 Stockholm, Sweden
E-mail: petter.holme@physics.umu.se, corresponding author

To understand the system-wide organization of metabolism, different lines of study have devised different categorizations of metabolites. The relationship and difference between categories can provide new insights for a more detailed description of the organization of metabolism. In this study, we investigate the relative organization of three categorizations of metabolites — pathways, subcellular localizations and network clusters, by block-model techniques borrowed from social-network studies and further characterize the categories from topological point of view. The picture of the metabolism we obtain is that of peripheral modules, characterized both by being dense network clusters and localized to organelles, connected by a central, highly connected core. Pathways typically run through several network clusters and localizations, connecting them laterally. The strong overlap between organelles and network clusters suggest that these are natural "modules" — relatively independent sub-systems. The different categorizations divide the core metabolism differently suggesting that this, if possible, should be not be treated as a module on par with the organelles. Although overlapping more than chance, none of the pathways correspond very closely to a network cluster or localization. This, we believe, highlights the benefits of different orthogonal classifications and future experimental categorizations based on simple principles.

With the advent of databases attempting to record the entire biochemical reaction systems of different organisms [1–7], a host of new research questions became accessible to researchers on metabolism. One did no longer have to restrict the research to certain subsystems such as the citric acid cycle or glycolysis, but could study system-wide organization of biochemical processes. Early studies in this vein found e.g. a skewed degree distribution [8], hierarchical modular organization [9–12], and a well-defined core and a modular periphery [13] (see Refs. [14] and [15] for reviews). Furthermore, these organization patterns are to some extent correlated with functionality [8–12] and the evolution [13,16,17] of metabolism system. One way of characterizing the large-scale structure of the metabolome is to divide the metabolites into categories, capturing some role or function (in a general sense) of the compound. These categories can be defined from the topology of metabolic networks or other types of information. A feature differentiating eukaryotic from prokaryotic cells is the presence of internal membrane-bound structures called organelles, such as nucleus, mitochondrion, and lysosome [18]. The subcellular compartmentalization by these organelles aggregates enzymes and substrates into spatially isolated localizations, and can therefore regulate of different metabolic processes [19,20]. Some algorithms have been proposed to predict the subcellular localization of proteins or metabolic enzymes [21,22]. However, researchers have not studied the organizational features of these units quantitatively much before. Recently, this has been possible thanks to two databases that include information on subcellular localization [5,6]. We use data from the BiGG database on the human metabolism [6].

In this study we investigate how three ways of categorization — pathways, localization and network clusters — are interrelated, and what their relationship can tell us about the system-wide organization of metabolism. The raw data for this study includes the lists of catalyzed reactions, localizations of metabolites and annotated pathways. We take an approach from social network theory and construct *block models* of the metabolites — networks where the nodes are the categories and two nodes are linked if they share a metabolite (or, in the case of network clusters, if they have an edge between them). This method is complemented by a topological analysis of a metabolic network using methods from the study of protein networks [23] (also inspired by social network analysis).

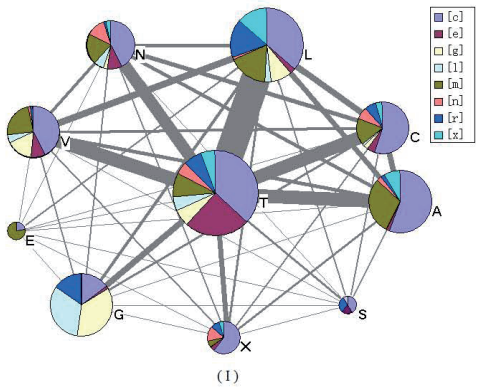
(I)

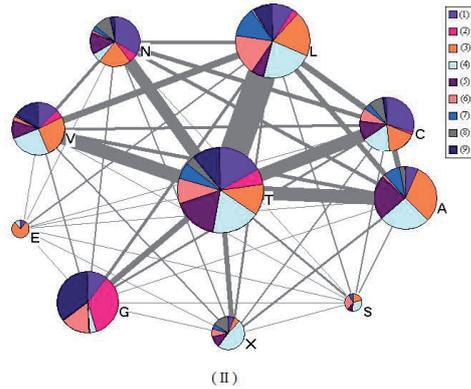
(II)

**Figure 1** — *Block-model representation of the pathways with cartographic overlays of subcellular localization I and network clusters II.* The sizes of the circles are proportional to the number of metabolites in each pathway. The width is proportional to the number of metabolites in common between the nodes. The initials of the pathways and localizations are given in the text.

## RESULTS AND DISCUSSION

**Relation between the categorizations — pathways, subcellular localizations and network clusters**

From the reaction system, we derive a substrate–product graph where the nodes represent metabolites and there is an edge between two metabolites if they occur in the same reaction and one is the product and the other a substrate of that reaction [14]. The resulting network consists of 2771 nodes and 9451 edges. From this network we identify *network clusters*, by the simulated-annealing algorithm of Ref. [10]. A network cluster identified by this algorithm is a region of the network that is more strongly coupled within than it to other clusters. Our second class is the *localization* (or *subcellular compartment*) of the metabolites, i.e. where in the cell a metabolite is occurring in a substantial amount. BiGG defines in total eight categories of this categorization. Our third class is the annotated *pathways* of the BiGG database. There is no general, widely accepted definition of a pathway. One common view is to start from one molecule (or a class of molecules) and define a pathway as the metabolic subsystem synthesizing or degrading these molecules [11]. But sometimes divisions are called pathways because of some common role of the metabolites. BiGG uses the ten pathways from the KEGG database, which contains divisions according to both these mentioned principles. Network clusters, localization and pathways are different ways of categorizing the metabolites, representing different traits of the metabolites; we will henceforth call them just *categorizations*. The categorizations are, we assume, neither completely overlapping (after all, there would be no need for different categorizations if they were), nor completely independent.

We start by analyzing the block-model networks [24] of our three categorizations. We construct weighted networks where a node represents a category of a categorization and the weight between two nodes is the number of metabolites in common between the two categories (for pathway block models), or the number of reactions between the two categories (for block models of subcellular localizations and network clusters). To picture the overlap between the different categorizations, we plot a pie chart per node of the relative number of metabolites of different categories belonging to one categorization (c.f. the cartographic representation of Ref. [10]). In this way, we can compare the categorization pairwise in six plots, one for each (ordered) pair plotted in Figs. 1–3.

In Figure 1 we plot the block-model representation of pathways and the cartographic representation of the overlap between pathways and the other two categorizations. The block-model network is a complete graph, showing that the pathways are all connected — material and information are exchanged between these metabolic subsystems. The transport (T) pathway has most interactions with the other pathways — implying, natural perhaps, that the transportation between distinct compartments plays an important role in cell metabolism linking other pathways together. Moreover, it can be seen that the sub-clique consisting of the classes amino-acid metabolism (A), carbohydrate metabolism (C), lipid metabolism (L), nucleotide metabolism (N), metabolism of cofactors and vitamins (V) and transport have many more interactions with each other. Except V these are all pathways of primary

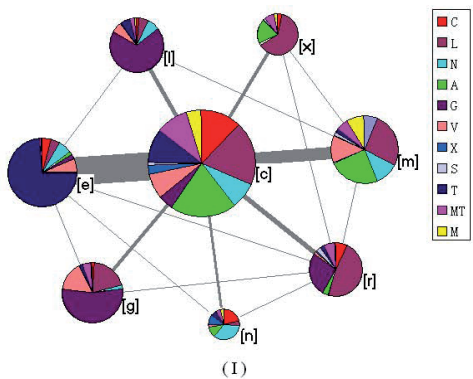
(I)

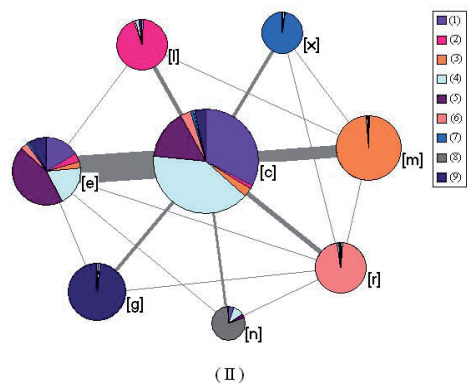
(II)

**Figure 2** — *Cartographic representations of the block-model graph of subcellular localizations.* The size of a circle is proportional to the number of metabolites in this category. The thickness of an edge is proportional to the number of reactions between the two corresponding categories. I Pathway cartography, where pie charts show the fraction of the pathways in the respective localization. II Topological cartography, where pie charts show the fraction of the network clusters in different subcellular compartments.

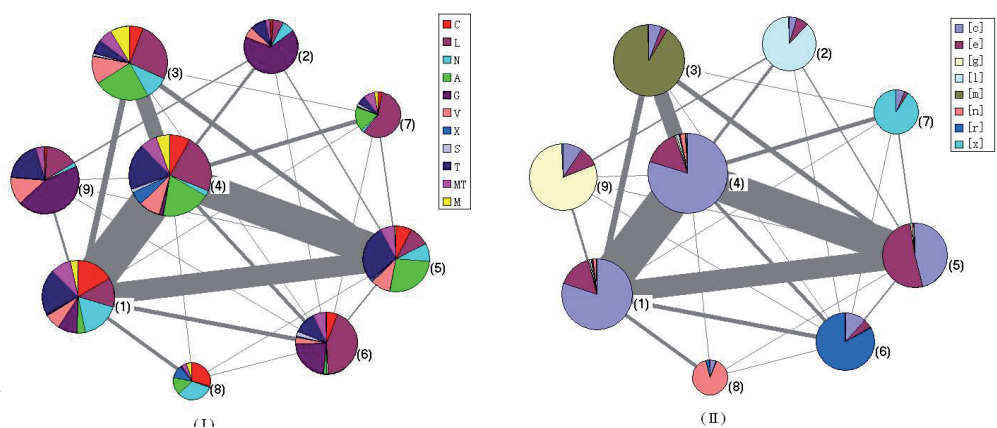

**Figure 3** — *Cartographic representations of the block-model with respect to network clusters.* Each circle represents a network cluster, while the edges reflect the connections between clusters. The size of a circle is proportional to the number of metabolites in this cluster, and the width of an edge is proportional to the number of links between the two corresponding clusters. The modularity metric [10] of this decomposition is 0.676. I Pathway cartography, in which the pie charts show the fraction of the pathways in that network cluster. II Localization cartography, in which the pie charts show the fraction of the subcellular compartments in that network cluster.

metabolism performing housekeeping functions. In contrast, the pathways of energy metabolism (E), glycan biosynthesis and metabolism (G), xenobiotics biodegradation and metabolism (X), and biosynthesis of secondary metabolites (S) have only thin links between themselves and other categories. Compared to the housekeeping pathways these carry out more specific functions. This linkage suggests that the pathways are also organized in a core-periphery pattern, like the substrate–product graph representation of human metabolic network [13].

In the cartographic plots of Figure 1-I and II we can see the distributions of subcellular localization and network clusters among the pathways. These graphs show the positions in the cell and the metabolic network of the pathways. All pathways are distributed over several subcellular compartments or network clusters, indicating that none of the pathways is restricted to a single place, neither to a cellular compartment, nor to a dense subnetwork. Especially, the primary metabolism pathways, A, C, L, N, and the transportation pathway have connections to almost all other subcellular compartments and network clusters, while the other metabolic functions are performed more localized. For instance, energy metabolism takes place only in cytoplasm and mitochondrion, or in network clusters 1, 3 and 4.

In Figure 2, we plot the block-model network of the subcellular-localization categorization. From its central position we can understand cytoplasm as a hub for material exchange and communication between different subcellular compartments. Actually, six of the eight compartments: mitochondrion (m), nucleus (n), Golgi apparatus (g), endoplasmic reticulum (r), lysosome (l), and peroxisome (x), are cellular organelles insulated from the cytoplasm by membrane envelopes. These organelles are most of the time physically separated and interact through the cytoplasm — one mode of interaction being the exchange of metabolites. Hence it is not surprising that the thickest line (strongest connection) between two location nodes is between cytoplasm (c) and extracellular space (e), reflecting the many ways material can be transportation across the cell boundary.

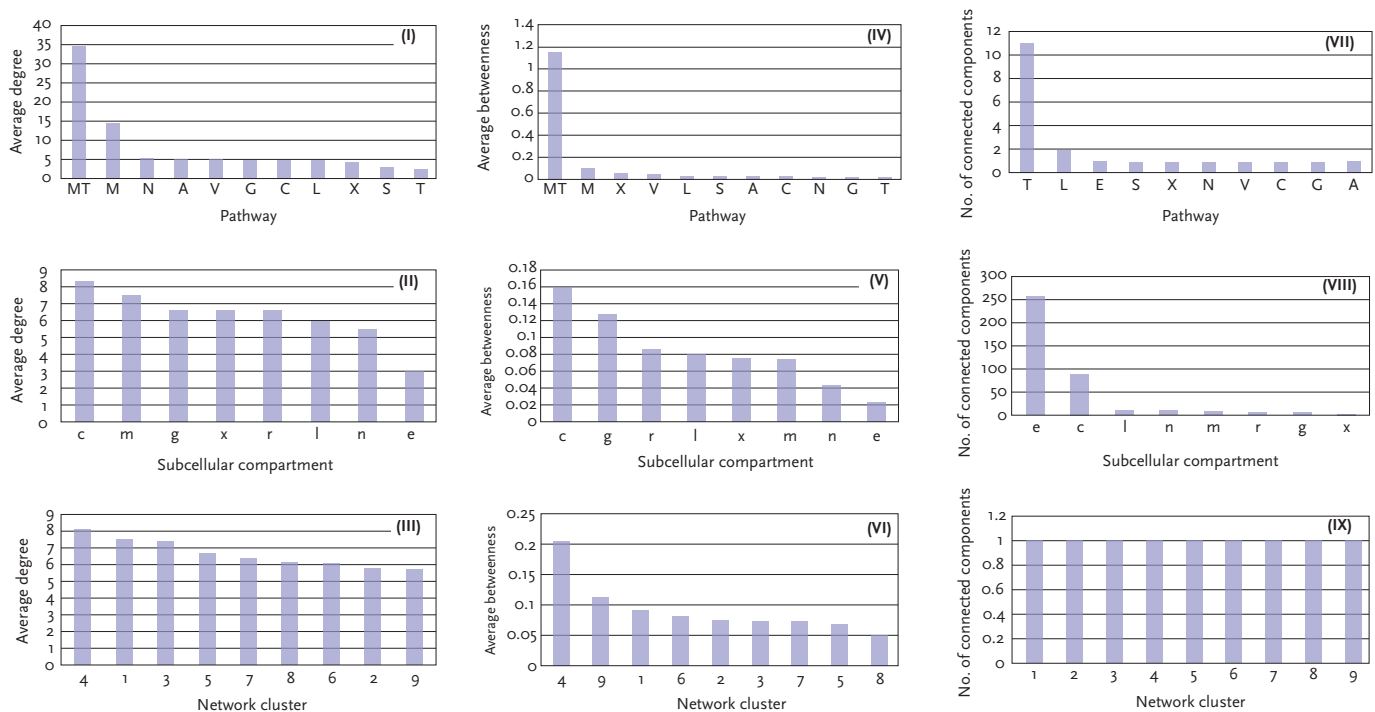

**Figure 4** — *Network structures of the categories.* Panels I–III show the average degree, IV–VI display the average betweenness and VII–IX the number of connected clusters. I, IV and VII are represent the pathways, II, V and VIII shows data for the subcellular compartments and III, VI, IX for the network clusters.

Figure 2-I shows the overlap between subcellular compartments and pathways. The pathways are not evenly distributed in the localization classes, suggesting that enzymes are segregated by the intracellular membrane systems. Five of the compartments — Golgi apparatus, endoplasmic reticulum, lysosome, peroxisome, and extracellular space — are dominated by one pathway. For instance, Golgi apparatus and lysosome are the main locations for glycan biosynthesis and metabolism; peroxisome and endoplasmic reticulum are the most important organelles for lipid metabolism; the transportation pathway dominates extracellular space. These observations are well-known in cell biology [18] — the Golgi apparatus is where for sorting and modification of proteins, including protein glycosylation; the lysosome breaks down large molecules such as proteins or glycans; the major function of peroxisome is to decompose fatty acids; and the endoplasmic reticulum mainly synthesizes lipids and steroids. Such aggregation of different enzymes in distinct spatial areas reduces the cross talk between metabolic pathways. Conversely, the other three compartments — the cytoplasm, nucleus and mitochondrion — are highly heterogeneous in terms of annotated pathways. This reflects the generality of the biological processes taking place in these localizations. The cytoplasm is the medium for the interaction between organelles, and also the compartment with the largest number of reactions; the nucleus is the source of genetic control of metabolism; and the mitochondrion generates much of the energy supply that drives the metabolic pathways.

As seen in Figure 2-II, the six localization categories of organelles correspond to one network cluster, respectively, suggesting a dense intra-organelle and sparse inter-organelle linkage of metabolites in the metabolic network. In contrast, the cytoplasm and extracellular space are not organelles but communication media of organelles and cells, respectively. Metabolites in these compartments are needed as in and output, not only for the organelles, but also for the relatively independent subnetworks forming our network clusters.

Figure 3 depicts the linkages among network clusters, which we obtained by the method presented in Ref. [10]. The structure of the block-model networks has the same type of a core-periphery organization observed in Ref. [13]. Clusters 1, 3, 4 and 5 are connected by many reactions, thus forming a core of the block-model network displayed in Fig. 3 and in the full metabolic network.

Unlike the pathways and subcellular compartments, network clusters are identified without any prior biological background knowledge (other than that the network itself is partly inferred from text mining). The usual interpretation of a network cluster is that it is a relatively independent subnetwork [25]; the same can be said of pathways, but the focus is more on the in- or output (how substances are synthesized or degraded). As seen in Fig. 3-I and II there is an overlap between network clusters and both pathways and subcellular localizations. Three of the four core clusters are mixtures of metabolites in cytoplasm and extracellular space, reflecting the central role of cytoplasm in cellular metabolism, and many opportunities of material exchange between the cell and its surroundings. The other clusters, including the five more peripheral categories and one core category, are dominant by metabolites from a single organelle, respectively, suggesting a high extent of overlap between the topological and localization categories. This feature implies that each organelle respectively defines a compact region in the metabolic network. As a relatively independent organelle, mitochondrion may have more complex functions and more interactions with the cytoplasm than the others. Projecting to topology, its corresponding network cluster, 3, is similarly a core cluster. From functional point of view, the core clusters are multi-functional categories in which multiple pathways are almost evenly distributed, whereas the peripheral categories exhibit to own a major function, for instance, glycan

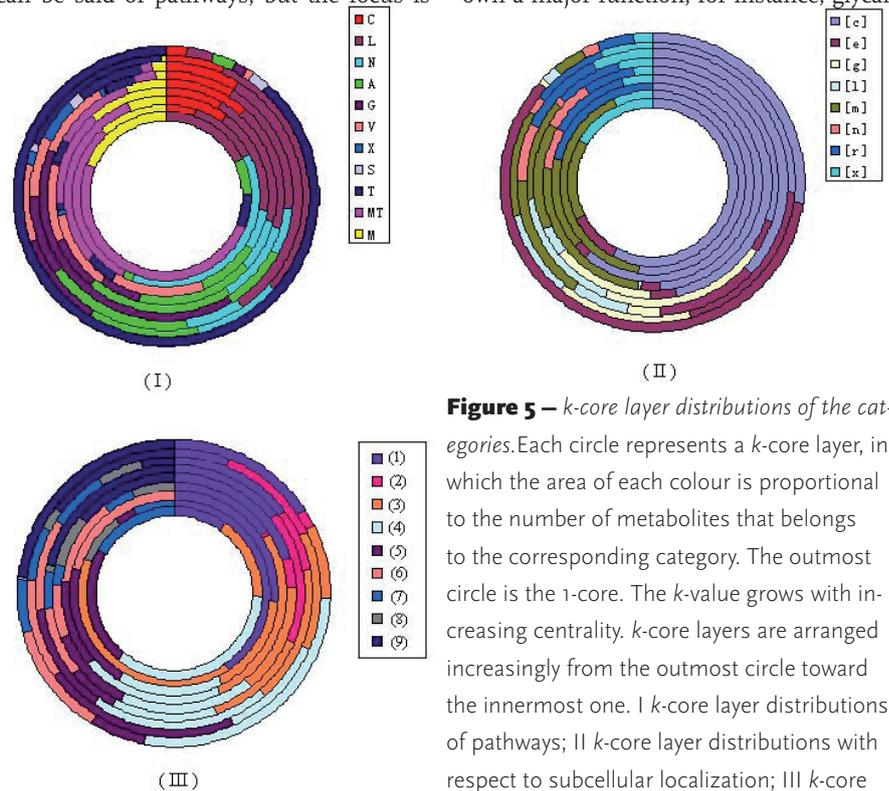

**Figure 5** — *k-core layer distributions of the categories.* Each circle represents a *k*-core layer, in which the area of each colour is proportional to the number of metabolites that belongs to the corresponding category. The outmost circle is the 1-core. The *k*-value grows with increasing centrality. *k*-core layers are arranged increasingly from the outmost circle toward the innermost one. I *k*-core layer distributions of pathways; II *k*-core layer distributions with respect to subcellular localization; III *k*-core layer distributions of network clusters.

biosynthesis and metabolism for 2 and 9, and lipid metabolism for 6 and 7.

**Quantitative difference between categorizations**

The cartographic block-model plots studied above give a very detailed picture of the relation between our three categorizations. In this section we try to condense this information to quantitatively answer how similar the three categorizations are. The similarity measure we use, $v$ [26], is (unlike the cartographic plots) symmetric. Its maximal value is one while zero represents neutrality. For localization and network clusters we obtain $v = 0.89 \pm 0.06$, for network clusters and pathways we get $v = 0.20 \pm 0.01$, and finally pathways versus localization gives $v = 0.21 \pm 0.02$. All values are well above zero (z-scores of $v$ are about 20), meaning that the categorizations are positively correlated, and thus to some extent they reflect similar aspects of biological organization. On the other hand, there are differences; so they do also measure different organizational traits. The subcellular localization and network clusters are overlapping most of the three pairs of functional categories. This effect seems to stem from the strong overlap between organelles and the peripheral network clusters (as seen in Figs. 2-II and 3-II). Since network clusters are thought of representing an organizational independence, and organelles are morphologically individual units, this seems natural.

**Network structure of the categories of the three categorizations**

So far we have studied the connections between the different categories and the overlap of categories from our three different categorizations. In this section we compare the structure of the individual categories in more detail.

From the early observations of broad and skewed degree distributions [8], we first examine the average degree for each category of the three categorizations, see Figure 4-I–III. For categorization according to different pathways, the hub metabolites are gathered in the categories "Multiple Functions and Transport" (MT) and "Multiple Functions" (M) (see Methods part for definition). These categories constitute only of 6% and 4% of the metabolites but are essential for keeping the network connected [8,27]. In contrast, the metabolites appearing in T (the pure transport category) have the lowest extent of metabolic interactions with others. One explanation of this is that MT does not include metabolites localized in extracellular space, whereas most metabolites of T are localized in extracellular space (66%) and cytoplasm (25%). The MT metabolites seem more involved in moving metabolites across intracellular membrane boundaries than the T metabolites that is more specialized in transport across cell walls. The corresponding study for the different subcellular compartments (seen in Fig. 4-II) indicates that metabolites in cytoplasm and mitochondrion have more interactions with others, and those in extracellular space have the lowest average degree. In Fig. 4-III) we see that for the network-cluster categorization the high-degree nodes are primarily located in the core clusters.

The node degree measures the local importance of a node. To get a more global view about the position of metabolites in the network, we also measure the betweenness centrality. The betweenness of a node is proportional to the number of shortest paths between pairs of nodes. Assuming metabolic processes preferably occur via short paths (which is not true for all processes [28]), betweenness should be a better indicator than degree for global centrality and importance. In particular, the betweenness (as the name suggests) is high for nodes connecting different network clusters. Figure 4-IV through VI shows the average betweenness for the different categorizations — pathways, subcellular compartments and network clusters, respectively. Metabolites in the MT pathway, localizations c and g, and network cluster 4 have high betweenness. One example showing that the global information of betweenness is more informative than the degree is that the Golgi apparatus is the organelle with highest betweenness, perhaps its central function of packaging macro molecules for secretion. For the network clusters, as can be guessed from Fig. 3, 4 is the cluster with highest average betweenness of the metabolites.

In Fig. 4-VII–IX, we investigate the number of connected components of the subnetworks of nodes of the same category. For the pathway categorization (Fig. 4-VII), we note that subnetwork defined by the transportation pathway is broken into over 10 isolated clusters (note that this pathway is also the sparsest, see Fig 4-I, sparse networks are naturally more prone to be disconnected). This highlights a difference between pathways and network clusters — pathways need not be independent units (an often quoted definition of "module") by construction. The energy metabolism pathway is connected (and also the second most densely linked pathway). Many metabolites of this pathway are small molecules normally labeled as carriers for transferring electrons or certain functional groups, such as ATP, NADH and $H_2O$. Such "currency metabolites" have often many more links than regular metabolites, explaining the density and connectedness of this pathway. Figure 4-VIII shows that the extracellular space compartment has a very fragmented network, even more than the transport pathway. The other non-organelle compartment, cytoplasm, is also disconnected. The organelle compartments, on the other hand, are connected. In sum, metabolites localized to the extracellular space and cytoplasm act as links between the more independent metabolic subnetworks of the organelles. Since our network clustering algorithm is designed to find densely connected regions it is no wonder that the network clusters are all connected (Fig. 4-IX).

**Network positions of the categories monitored by *k*-core decomposition**

A so-called *k*-core decomposition is a way to visualize both how connected neighborhoods of nodes are and their centrality [23]. Stated briefly, it is obtained by iteratively deleting low-degree nodes to achieve a sequence of *k*-cores (maximal subgraphs with minimal degree *k*, see the Methods section), so that following the decomposition is like zooming in to the more central and more interconnected parts of the network. For the substrate–product network we study there has a 9-core but not a 10-core. In Figure 5, we investigate how the categories are distributed in the $k$-cores. Most nodes placed in the outmost core layer are from the T pathway (Figure 5-I) or subcellular compartments e and c (Figure 5-II), suggesting that a large part of metabolites functioning in pure transport or situating in non-organelle compartments are input or output metabolites only produced or consumed in metabolism. The categories can be roughly partitioned into two types according to their node distributions in the $k$-core layers. Type 1 includes categories whose nodes appear in almost every $k$-core layer or in relatively inner $k$-core layers, such as pathways C, L, N, A, MT, and M (Figure 5-I); localizations c, e, m, and x (Figure 5-II); and network clusters 1, 3, 4 and 5 (Figure 5-III). Categories of this type tend to be, with respect to pathways, primary metabolism or multiple-function categories; with respect to localization, material exchange media or organelles with relatively independent function; and with respect to network topology, core clusters. The other type includes categories whose nodes are concentrated to the outer $k$-core layers, such as pathways S, X, G, T (Figure 5-I); localizations g, l, n (Figure 5-II); and network clusters 2 and 9 (Figure 5-III). These categories are more related to input and output to the system, and more specialized processes.

## CONCLUSIONS

To understand a large system such as the metabolism one need to simplify and categorize its components. In this paper we have investigated three ways of doing this — grouping metabolites according to pathways, localization, and network clusters. Our main method is inspired by the block modeling of social network studies and cartographic plots of Ref. [10]. The compartmentalization is clearly organized into a core of extracellular media and cytoplasm, and a periphery of organelles. There are peripheral network clusters overlapping almost completely with the organelle categories. The cores of both categorizations, according to subcellular compartments and network clusters, are more highly connected than the peripheries. These observations are together painting a picture of a global organization of the metabolites into a core (identified by both average degree, betweenness and k-core decomposition) and a periphery of modules — relatively independent units — corresponding to the organelles. Some studies [19,20] point out compartmentalization as an important structure for regulating metabolism, other studies say the same thing concerning network clusters [10–13,29]; our results suggest that these ideas can be two sides of the same coin — peripheral, spatial compartments of the cell have corresponding network clusters. The central localizations are fragmented — for atoms of one molecule to be converted to another, chains of reactions leaving the compartment is needed. In other words, they are not independent sub-network, an thus not modules. The pathways do overlap with both subcellular localizations and network clusters more than what is expected by chance, but much less than the overlap between localizations and network clusters. This, we believe, reflects the different philosophy behind pathways — many pathways are defined as the subnetworks participating in the biosynthesis or degradation of some class of molecules; and as such, pathways can pass different localizations and network clusters. Other pathways, at least the transportation pathway, are defined via the physical function of the metabolites, and stand out by being low-connected and fragmented. To look forward, we believe there is much qualitative knowledge to be gleaned by investigating the relative organization of different categorizations, rather than the individual categories; but for this method to be fully successful all categories of a categorization need to be identified by the same quantitative criteria.

## METHODS

### Data description

Our raw data was obtained from the BiGG [6] database of metabolic networks. This data was manually reconstructed component-by-component based on genomic and bibliomic data. The BiGG database includes a list of 3311 reactions occurring in the following eight subcellular compartments (mentioned in the Results section). The pathway annotations (of the ten, above mentioned, pathways) originated from the KEGG database [1] (www.genome.jp) where reactions are labeled by the pathway. We assign the pathways of a reaction to its participating metabolites. Every metabolite is thus associated with at least one pathway. When we analyze the network topological features of the nodes, we need each metabolite to belong to only one pathway. To achieve this, we added two pathway categories for metabolites in multiple pathways according to the following scheme:

- For metabolites belonging to two pathways including transport, we assign them to the other pathway class than transport.
- For metabolites that belong to at least two pathways not including transport, assign them to the "Multiple Functions" pathway (M).
- For metabolites that belong to at least three pathways including transport, assign them to the "Multiple Functions and Transport" pathway (MT).

### Network construction

**Including compartment information in the substrate–product network**: There are many kinds of graph representations of metabolism [14,30]. In this study, all of the reactions in BiGG database were used to reconstruct human metabolic network we study. In this network, one node is a metabolite in a specific subcellular compartment. For example, according to BiGG, glucose–6–phosphate is localized to both the compartments c and r giving two nodes in our network.

**Block-model network of categories:** Block modeling is a general way of structuring and simplifying large-scale organization commonly used in social science [24]. In this methodology one construct a network of classes of nodes that can be linked in various ways. This higher-order network can then be analyzed with general network methods. In our study, the nodes of the block models are the categorizations of the pathway, subcellular-localization and network-cluster categorizations, the meaning of the edges are defined in the captions of Figs. 1–3.

**Clustering the substrate–product network:** To achieve the network clusters we use the same method and parameter values as in [10]. The general philosophy of this method is to maximize a measure of modularity of partitions of a network [31]. Simulate annealing, the optimization method we use, mimics Monte Carlo simulations of statistical mechanics. By allowing some disorder the algorithm avoids getting stuck in local minima. We compared the results with a more specialized algorithm [32], but the simulated annealing algorithm could (somewhat surprisingly) find partitions with larger modularity than this method.

**Matching between different categorizations**

For measuring the similarity between different categorizations, we use a method described in Ref. [26]. Consider two categorizations $X$ and $Y$ (for example be subcellular localization and pathways) and assume each metabolite is associated with a subset of the categories of $X$ and $Y$. Let $\phi_X(x)$ denote the fraction of metabolites in category $x \in X$, and define $\phi_Y(y)$ correspondingly. Let $\phi_{XY}(x,y)$ denote the joint frequency of $x$ and $y$, i.e. the fraction of vertices that are categorized both as $x \in X$ and $y \in Y$. In a random distribution of functions the expectation value of $\phi_{XY}(x,y)$ is $\phi_X(x)\phi_Y(y)$, but if the categories of different categorizations are overlapping, then some $\phi_{XY}(x,y)$, the ones that overlap, will be larger than $\phi_X(x)\phi_Y(y)$, while for the others $\phi_{XY}(x,y)$ will be lower than $\phi_X(x)\phi_Y(y)$ (since $\phi_{XY}(x,y)$ need to add up to unity and much is spent on the overlapping categories). Thus, both overlapping and not overlapping categories will contribute to $|\phi_{XY}(x,y) - \phi_X(x)\phi_Y(y)|$ and a prototypical overlap score is

$$\mu = \sum_{x \in X}\sum_{y \in Y} |\phi_{XY}(x,y) - \phi_X(x)\phi_Y(y)| \qquad (1)$$

This quantity is, however, affected by finite sizes so that is hard to estimate if a $\mu$-value is larger or smaller than expected. Instead, as our operational matching measure we use $\nu$, the Coleman coefficient of $\mu$:

$$\nu = \frac{\mu - \bar{\mu}}{\mu^* - \bar{\mu}} \qquad (2)$$

where $\bar{\mu}$ is the average $\mu$-value over a randomized assignment of categories (with the only constraint that the size of the categories are the same as in the real data and no category can be assigned twice to the same metabolite), and $\mu^*$ is the maximal value of $\mu$ in the same ensemble.

**Betweenness**

Betweenness is proportional to the number of shortest paths between other node pairs that pass a node. More technically, let $\sigma(s,t)$ denote the number of shortest paths from $s$ to $t$ and $\sigma_v(s,t)$ denote the number of shortest paths from $s$ to $t$ passing $v$, then the betweenness is given by [24]:

$$C_B(v) = \sum_{s \neq t} \frac{\sigma(s,t)}{\sigma_v(s,t)} \qquad (3)$$

**$k$-core and $k$-core layer**

The $k$-core of a graph is the maximal subgraph such that all its nodes has at least $k$ links within the subgraph [23,33]. The $k$-core layer $l_k$ is defined as the set of nodes that belong to $k$-core but not to $k+1$-core, i.e., $k$-core is the union of $k+1$-core and $k$-core layer. A $k$-core subgraph of a graph can be generated by recursively deleting the vertices from the graph whose present degree is less than $k$. This process can be iterated to gradually zoom into the more connected parts of the network. The higher-level core corresponds to more densely connected part of the network. See Figure 6 for an explanation.

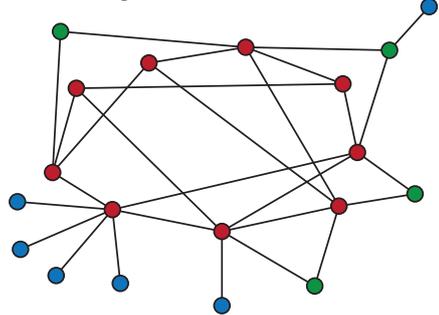

**Figure 5** — *k-core layer distributions of the categories.* Each circle represents a $k$-core layer, in which the area of each colour is proportional to the number of metabolites that belongs to the corresponding category. The outmost circle is the 1-core. The $k$-value grows with increasing centrality. $k$-core layers are arranged increasingly from the outmost circle toward the innermost one. I $k$-core layer distributions of pathways; II $k$-core layer distributions with respect to subcellular localization; III $k$-core layer distributions of network clusters.

## AUTHORS' CONTRIBUTIONS

JZ and PH conceived of the study, designed the analysis, implemented the analysis and wrote the manuscript together.

## ACKNOWLEDGEMENTS

PH acknowledges support from the Swedish Foundation for Strategic Research.